\documentclass[a4paper, biblatex,]{jacow}

\newcommand{\eg}{\emph{e.g.}}
\newcommand{\Eg}{\emph{E.g.}}

\newcommand{\incl}{\emph{incl. }}

\begin{document}

\title{Open, to What End?\\A Capability-Theoretic Perspective on Open Search}

\author{
Nicola Neophytou\thanks{neophytounicola@gmail.com}, Independent Researcher, Tiohtià:ke / Montréal, Canada \\
Bhaskar Mitra\thanks{bhaskar.mitra@acm.org}, Independent Researcher, Tiohtià:ke / Montréal, Canada \\
}

\maketitle

\begin{abstract}
The hegemony of control over our search platforms by a few large corporations raises justifiable concerns, particularly in light of emerging geopolitical tensions and growing instances of ideological imposition by authoritarian actors to manipulate public opinion.
Recent movement for promote open search has emerged in response. 
This follows from past and ongoing push for openness to challenge corporate oligopolies (\eg, open source and open AI models) which have seen significant ongoing negotiations and renegotiations to establish standards around what constitutes being \textit{open}.
These tensions have hindered these movements from effectively challenging power, in turn allowing powerful corporations to neutralize or co-opt these movements to further entrench their dominance.
We argue that the push for open search will inevitably encounter similar conflicts, and should foreground these tensions to safefguard against similar challenges as these adjacent movements.
In particular, we argue that the concept of \textit{open} should be understood not with respect to what is being made open but through a capability-theoretic lens, in terms of the capabilities it affords to the actors the system is being opened to.
\end{abstract}
\section{Introduction}
\label{sec:intro}
Search systems inherently mediate what information receives exposure, and in doing so shape our political opinions, what we choose to consume, and how we perceive our place in society.
The hegemony of control over our search platforms by a few large corporations has therefore triggered justifiable concerns, especially in light of emerging geopolitical tensions between the US and its allies~\cite{weissert2026trump, mancini2026canada} and increasing instances of ideological imposition by authoritarian state and private actors to manipulate public political opinion~\cite{noble2025president, lu2025we, kerr2025musk, taylor2025musk}.
These concerns are compounded by the increasing integration of generative AI models in information access, which are becoming increasingly effective in manipulating mass opinion~\cite{mitra2024sociotechnical}.
In response, there has been recent calls for prioritizing societal needs over corporate interests~\cite{mitra2025search} in search, counter-imagining hegemonic search as foundational infrastructure~\cite{mager2023european}, establishing democratic digital public sphere~\cite{grassmuck2021towards}, developing public interest search platforms~\cite{plote2025charter}, and safeguarding our information ecosystem from authoritarian and corporate capture~\cite{mitra2025emancipatory, mitra2026information}.
Each of these calls underscore the need for open search alternatives.

From source code to AI models, computing has a long tradition of pursuing openness as a mechanism to challenge capitalist and oligopolistic power.
These past and ongoing movements represent sites of conflict between corporations seeking industry dominance and the impacted communities exercising resistance.
Within these struggles, the standards of openness are constantly negotiated and renegotiated.
These tensions can be observed, for example, in the divergence of the so-claimed technological pragmatism of the open source movement from the sociopolitical aspirations underpinning the free software movement~\cite{stallman1998why}.
Also notable in this struggle are the strategies employed by large corporations to defend their oligopolistic power---\incl smearing~\cite{raymond2002microsoft}, sabotaging~\cite{economist2020deadly}, co-option~\cite{lund2020profiting}, and open-washing~\cite{curtis2012beware}.
These tensions have been renewed in recent years within the open AI movement, where the definition of ``open'' is under contestation~\cite{widder2023open, ftc2024open} and open-washing is rampant~\cite{widder2024open, liesenfeld2024rethinking}.
In response, several frameworks have been proposed to measure openness of AI models~\cite{solaiman2023gradient, white2024model, liesenfeld2024rethinking}, grading them based on what aspects of the model and its production process are being made open.
The open search movement is still in its infancy, yet we can observe renegotiations on the notion of openness in the process of its conception~\cite{mager2023european}.
We posit that a reexamination of the concept of ``open'' is necessary because of the shared tensions between open search and previous open movements.

Movements for openness---be it for source code, models, data, or systems---are generally motivated by specific social, political, economic, or technological aspirations.
However, the negotiations that follow often center \textit{what} is being made open rather than whether the act of opening brings us closer to our stated motivations.
This creates opportunity for powerful corporations to benefit reputationally from participating in the open ecosystem and extracting value from the movement to further entrench their dominance, without substantially satisfying any of the socioeconomic or political motivations of the open movement.
Even if the system is being made open with sincere intentions, the inordinate focus on what is being made open detracts the community from undertaking the critical examination of: \textit{to what end} does this opening contribute, and who benefits from it?

Towards this goal, we argue that openness should be understood and appraised with respect to the capabilities it affords to actors that the system is made open to.
We draw from Amartya Sen's Capability Approach (CA)~\cite{sen1999commodities} to reframe \textit{openness} centering desired well-beings, and corresponding capabilities actors must be granted to achieve them. 

\section{Why do we need open search?}
\label{sec:why}

No search platform is ``neutral'' as every decision about how to organize and surface information is saliently political.
Every stage of their design involves normative choices that determine what information is deemed relevant and whose perspectives are considered worthy of exposure.
Dominant search companies make these decisions behind closed-doors, sheltered from public scrutiny.
This lack of transparency and accountability raises concerns that search systems may inadvertently (or deliberately) foster ideologies that align with those of the platform operators, or the state actors who have sway over them.
\Eg, U.S. president Donald Trump's ``Preventing Woke AI in the Federal Government'' executive order imposes the Trump regime's own ideological agenda~\cite{noble2025president} on AI models, with implications for the search systems that incorporate them.
Given existing evidence that search results can be instrumental in influencing political opinion~\cite{epstein2015SEME} and emerging evidence on AI persuasion~\cite{mitra2024sociotechnical}, this raises serious concerns for how public opinion is vulnerable to manipulation by private corporations and authoritarian regimes.

The European Union (EU)'s attempts to rein in Big Tech's power through regulations have been met with intense lobbying and political maneuvering~\cite{schyns2023lobbying, gorwa2024platform, corporate2026article}.
Concurrently, rising geopolitical tensions involving the US~\cite{weissert2026trump, mancini2026canada} has attracted further attention to the vulnerabilities of a digital ecosystem in which many countries rely heavily on the tools controlled by a few US-based firms.
In response, several countries, including Canada~\cite{djuric2025canada} and in the EU~\cite{fitzGerald2025digital}, are accelerating \textbf{digital sovereignty} mandates to enhance the autonomy of their national data and digital infrastructure.
These initiatives seek to reduce dependence on overseas tools by creating self-hosted and local alternatives, offering industry for local economies, and the opportunity to engrain a culture of compliance-by-design with local laws.
Projects like the Open Web Search (OWS)\footnote{\url{https://openwebsearch.eu/}}, and the related Open Web Index (OWI)\footnote{\url{https://openwebindex.eu/}} initiative, funded by the EU and ``based on European values''~\cite{CERN2025european}, are examples of how open and publicly governed search alternatives are emerging to address concerns around digital sovereignty and promote global competition.

An open search ecosystem may also encourage investments in niche markets and \textbf{addressing historically underserved linguistic, regional, and sociocultural needs} of communities, filling important gaps where the current dominant search platforms do poorly.
This includes robust information access in and for under-served languages, ensuring appropriate representation of historically marginalized peoples and cultures, indexing and retrieving regionally relevant and community-specific knowledge and cultural artifacts, and addressing local and community-specific needs.

The development of open search also lowers the bar to entry for new search providers to compete by leveraging shared infrastructure without needing to build from scratch.
This has the potential to encourage a faster pace of \textbf{innovation through market competition}.
One can point to historic cases of innovation under open internet standards to support the movement for shareable and co-optable infrastructure~\cite{internet2015policy}.
This is particularly relevant in light of US Big Tech monopolies stifling competitors both in the US~\cite{reuters2024monopoly, DOJ2025antitrust} and globally---\eg, in the EU~\cite{Milmo2024EU, Ulea2025EU}.
However, in the realm of AI, in spite of a burgeoning ecosystem of open models, smaller companies have struggled to be competitive, and the economic benefits of AI still flow exclusively through the major AI monopolies~\cite{widder2024why}.
The open search movement must therefore consider that the reality of achieving real competition goes beyond just creating an open ecosystem, and must circumvent the levers of power that Big Tech instrumentalizes to maintain their dominance.

Open search also supports \textbf{promoting open science and rigor in information access research}, echoing the ethos of scientific reproducibility and lowering the barrier between academic research and production systems.
Open search in this context may enable researchers to conduct their experiments under more realistic system settings, provide access to search logs and live traffic that academic researchers typically do not have access to, and make it convenient for the community to reproduce and build on each other's work.

Furthermore, an open search ecosystem also presents an opportunity to progress towards recent calls~\cite{mitra2025search, mitra2025emancipatory, trippas2025report} for \textbf{realizing information access as a force for justice, emancipation, and democracy}.
Profit motives have historically been the key driving force for proprietary platforms.
\Eg, in the social media space, Meta has repeatedly evaded the responsibility to reduce misinformation and radicalization on its platforms~\cite{facebookfiles}, echoing how privately-owned platforms prioritize engagement and growth at the cost of information integrity and societal good.
An open search ecosystem is an opportunity for us to deliberately consider how we can build information access platforms centering our societal and democratic needs, instead of being reactionary to potential social harms of emerging information technologies~\cite{mitra2025search}.
This requires radical rethinking of the design of our search systems and developing effective mechanisms for democratic governance of our platforms since we cannot simply rely on state regulations to uphold our ethical values~\cite{mitra2026information}.
Compliance with regulations does not necessarily imply that societal concerns are wholly accounted for, and does not ensure genuine emancipatory outcomes. 

Next, we present a capability-theoretic perspective on open search to formalize the connection between these motivations and the capabilities required to attain them.

\section{Rethinking Search Openness }
\label{sec:open}

We present an alternative perspective on openness in search, shifting the focus from assessing transparency and access to individual stack components to instead evaluating openness by what \textit{capabilities} are enabled by opening up the system.
This theoretical move uncovers two key insights currently underrepresented in the discourse around open search; first, openness is a means to an end and should be appraised with respect to the affordances that result from it, and second, the capabilities that result from system openness are not universal and instead depend on the circumstances of actors that the system is purportedly being opened to.
We draw from Amartya Sen's Capability Approach~\cite{sen1999commodities} from welfare economics to reframe openness for search.
While our focus is on open search, we believe this reframing also speaks to the shared underlying tensions we observe across other movements for open source, data, models, and science.
We posit that the open search movement would benefit from adopting the capability-theoretic lens to reexamine its motivations and how purported strategies of \textit{opening} the system is supposed to help move us towards those goals.

\subsection{The Capability Approach}
CA outlines a philosophical framework for evaluating a person's well-being in terms of their capabilities and functionings~\cite{sen1999commodities}.
Sen defines a \textit{functioning} as an ``achievement of a person: what he or she manages to do or be'' which may contribute towards their subjective \textit{well-being}.
A \textit{capability} then reflects a person's ``ability to achieve a given functioning''~\cite{saith2001capabilities} given the \textit{commodities} available to them.
\begin{equation*}
    \textit{Commidity} \rightarrow \underset{\text{(to function)}}{\textit{Capability}} \rightarrow \textit{Functioning} \rightarrow \textit{Well-being}
\end{equation*}
CA emphasizes that commodities alone do not result in desired functioning or state.
Importantly, the ability to achieve a functioning, given a commodity, diverges among people with varying circumstances.
For example, individuals may require different resources to move around a building depending on their mobility needs.
If everyone is presented with a staircase (\textit{commodity}), their ability to move across floors (\textit{functioning}) may not be the same.
We leverage CA to underpin the idea that openness in search should be appraised with respect to its ability to enable desired functionings, and in light of the divergence in capability that it affords to different actors.
This perspective is underrepresented in current open search discourse, and constitutes our primary contribution.

We begin by positioning the previously stated motivations for open search as desired \textit{well-beings} within our framework, specifically, achieving digital sovereignty, addressing historically underserved linguistic, regional, and sociocultural needs, encouraging innovation through market competition, promoting open science and rigor in information access research, and realizing information access as a force for justice, emancipation, and democracy.
We then track backwards to identify key corresponding \textit{functionings}---a set of ideal `beings and doings'---that should be afforded by open search to actors to move towards said well-beings, namely to
\begin{enumerate*}[label=(\roman*)]
    \item critique,
    \item copy, and
    \item co-opt and co-construct the search system.
\end{enumerate*}
Next, we enumerate different \textit{commodities} including code, data, models, tooling, compute, labor, and governance that open search may make available.
We conclude this section with a discussion of how open commodities afford different \textit{capabilities} to different actors.

\subsection{Well-beings}
The motivations for open search span a range of political, sociocultural, economic, scientific, and societal causes that we map to desired well-beings under CA.
These intrinsically different motivations make saliently different demands of openness from search systems.
For example, open search is not a strict prerequisite for \textbf{digital sovereignty} nor for \textbf{addressing historically underserved linguistic, regional, and sociocultural needs}.
There are several examples of  privately-owned search engines that focus on the needs specific to a language, country, or user demographic, \eg, Baidu (for Chinese), Yandex (for Russian), Naver (for Korean), Parsijoo (for Persian), and KidRex (for kids).
However, the path to digital sovereignty or meeting needs of local communities likely require challenging US Big Tech hegemony over search platforms and open search maybe critical to bring together industry, academia, government, and civil society to mount a more concerted offensive.
Open search is also important if the goal is not to just address the needs of well-resourced individual states or communities; but to develop replicable strategies and shareable artifacts for those who may lack the adequate resources to build for their own needs.

If the goal is to encourage more \textbf{innovation through market competition} then providing open artifacts under appropriate commercial-usage licensing would be valuable.
Commercial entities that leverage these artifacts have access to disparate level of resources of their own and therefore will draw different capabilities from the same commodities.
\Eg, if a high-volume data firehose was made available as part of open search, not all actors may have the necessary infrastructure to consume it.
Similarly, if the learned parameters of a massively large model was made open, not every actor will have the compute resources to deploy that model in production settings or finetune it for their specific application scenario.
The criticality of these considerations depends on how seriously we want to safeguard against the possible emergence of an oligopoly of actors leveraging these commodities more effectively due to having outsized access to external resources relative to their competitors.

For \textbf{promoting open science and rigor in information access research}, open search can enable information access researchers to conduct their studies and test their hypotheses under reproducible real-world deployment settings.
However, the additional overhead of working with large-scale production systems may also hinder researchers who may be more interested in access to a simpler prototype of the system that they can easily work with and not have to undergo a steep learning curve.
So, the openness that academics may desire from search system are likely different from someone who wants to leverage its openness with commercial intent.

Finally, if our objective is \textbf{realizing information access as a force for justice, emancipation, and democracy}, then it is important to ensure equity of capabilities that different communities can draw from the artifacts made open.
This is because marginalized communities, who are interested to leverage openness to critique and co-produce technology in support of their struggles, are also likely to be generally resource-poor.
Johnson~\cite{johnson2014open} makes a similar critique of the benefit of open data for democracy and justice that is often taken for granted; and instead argues that open data raises serious concerns of information justice.
Furthermore, epistemic justice~\cite{fricker2007epistemic} and representational justice~\cite{lewis2016vision} may not be achievable simply by opening the system to critique and to be copied for adaptation.
To challenge the hegemony of western epistemologies and the seeing of the world through the dominant Eurocentric, white, male gaze, we require marginalized communities to also be able to co-opt, deconstruct, and reshape the search system being made open~\cite{mitra2026information}.

\subsection{Functionings}
To achieve the aforementioned well-beings relevant to open search, we identify three key functionings within our capability-theoretic framework which we discuss next.

\subsubsection{Critiquing.}
Critiquing refers to the study of sociotechnical systems like search engines to identify potential risks and mechanisms of marginalization and other harms from the system.
The object of critique may include the system design (\eg, accessibility of user interface), the materials employed in the production of the system (\eg, datasets) and components of the technology stack (\eg, models), the quality control and evaluation protocols (\eg, metrics for decision-making), the institutional policies and guidelines for system development and operations (\eg, moderation policies and guidelines for data labeling), the organizational processes involved (\eg, red-teaming and system monitoring), the incentive structure that shapes the system (\eg, monetization strategy), and how these aspects interact in complex ways.\footnote{
\Eg, If the deployment of new ranking models depends on ad revenue from clicks, the system may be at risk of optimizing for the set of users generating the most revenue, who tend to belong to historically privileged demographics, at the cost of worse relevance for marginalized groups; alienating the latter and causing them to abandon the platform; in turn reducing their representation in search log datasets used for training ranking models making them further underserve marginalized peoples; while online success metrics like clickthrough rate falsely indicate higher average user satisfaction because of the exclusion of underserved users.
}
Critiquing maybe conducted by academics and civil society to safeguard against societal harms, or by regulatory bodies to ensure legal compliance (\eg, antitrust laws, consumer protection laws, data protection laws, and emerging AI regulations); and is emphasized as one of three practices in the emancipatory information access framework~\cite{mitra2025emancipatory}.

\subsubsection{Copying.}
Copying refers to building alternative systems by leveraging components of an open search.
The OWI is a relevant example of this.
Its goal is to support the development of a range of new search systems by permitting their use of the open web index it makes publicly available.
Such commodities can support the development of alternative systems and infrastructure to challenge the hegemony of big platforms and address unmet community needs in search.

While copying may be a possible path to address the needs of local communities not offered by dominant platforms, it cannot meaningfully mitigate the epistemic injustice and representational harms caused by said platforms.
Building separate, distinct search engines does not solve the issue of harmful content, e.g., the sexualization of women of color in search results~\cite{noble2018algorithms, urman2024foreign}, being proliferated on the dominant platforms if they maintain their monopolistic control over market share. 
Furthermore, it is also undesirable that every community needs to build a new system to mitigate the erasure of their knowledge or their harmful representation.
Even if that is practically and economically feasible, it would produce a range of different systems that neither meaningfully challenge dominant platforms on their own nor benefit from the work of others without massive global coordination. Evidently, there are limitations to the possible gains from being able to simply copy from and make alternatives to dominant systems, particularly when evaluating whether openness satisfies the goal for information access as a force for justice.
This leads us to consider the next functioning.

\subsubsection{Co-opting and co-constructing.}
The most radical functioning that open search could enable is to allow for communities to co-opt and co-construct the open search platform itself, which Mitra et al.~\cite{mitra2026information} argue is crucial if our goal is to realize information access as an emancipatory force.
In this proposal, communities modify the open search platform to ensure epistemic and representational justice for themselves, and are not restricted to just make derivative systems from it.
The platform in turn becomes a site upon which the efforts of several different communities can be put into meaningful arrangements to design information experiences that meet the needs of each.
For such arrangements to emerge, the open platform design must hold space for inter- and intra-community negotiations and put in place platform governance mechanisms to ensure specific outcomes, such as foregrounding justice, emancipation, and democracy.

\subsection{Commodities}
The commodities that open search systems cann make available include code, data, models, tooling, compute, labor, and governance.
Of these, code, data, models, and tooling constitute artifacts that has traditionally been considered for making open in computing.
In the pursuit of open AI models, Widder et al.~\cite{widder2024why} emphasize the need for also making open the compute and the labor necessary for AI production.
For open search, both the compute and the labor necessary for the development and running of search systems should likewise be subject to considerations for opening.

In this context, compute includes the hardware and computational resources necessary for indexing, model training, and online running of the retrieval pipeline; and the human labor necessary for search systems includes data labor for relevance labeling, content moderation, red-teaming, and other data annotations, as well as labor involved in system engineering and operations.
Making labor available as a commodity may include allowing critique of working conditions and compensation; providing access to pool of crowdworkers and engineers for derivative system development; and the option to contribute external workforce to the pool.
Additionally, we posit that open search should also make available the processes, tools, and other resources for platform governance.
Here, labor and governance constitute the \textit{social infrastructure} that are also candidates for opening.

\subsection{Capabilities}
What functionings are afforded by the act of making the system open is determined by a combination of \textit{what} commodities are made open in the process, \textit{how} they are made open, and the circumstances of \textit{who} they are made open to.
These factors interact in complex ways and are highly contextual to the specific functionings we want to achieve.

For example, making a search system available as a runnable application would support critique and copy only if the person critiquing or copying have access to necessary compute and data to run the system.
Alternatively, components of the system could be exposed as an application programming interface (API) which may be suitable for someone trying to leverage the component in their own system or critique the system outputs without necessitating significant new compute resources; but this is suboptimal if the goal is to scrutinize how the component functions for a system audit. 
Availability of corresponding documentations, and the quality thereof, may also be instrumental in determining the utility of open commodities.
It is also important to consider the granularity of access being granted.
For example, when making the code or runnable binaries for indexing available, do we also make available the same for document parsing?
Are there external packages or services that indexing employs that should also be made open?

We must also critically interrogate who the system is being made open to and for what permitted purposes.
These constraints can impede achieving desired functionings and in turn the desired goals of open search.
\Eg, search platforms may provide targeted access to selective auditors for regulatory compliance or to earn social goodwill, but prevent academia and civil society from meaningfully scrutinizing the system.
This may create false perceptions of openness that reputationally benefit platform owners without resulting in any meaningful public oversight, \eg,~\cite{solon2020while}.
It is crucial for us to also pay attention to restrictions that may be levied on open commodities.
\Eg, system components may be made available with non-commercial licensing that allow companies to reputationally benefit from having made their system open without endangering their competitive advantage nor contributing to improving market competition.
Similarly, copyleft licensing may deter some from adopting open commodities, but such licensing can also be beneficial for soliciting reciprocal openness~\cite{frantsvog2012all} and safeguarding open search from exploitation by dominant market players.

The CA lens foregrounds these tensions and encourages the open search community to engage in critical dialog and find basis of solidarity across diverse motivations.

\section{Conclusion}
\label{sec:conclusion}
We instrumentalize capability theory to provide a different vantage point from which to examine open search.
We illustrate how different desired states of well-being demands different functionings from opening the system.
The open search path for digital sovereignty, for example, need not run parallel to the path for social justice and emancipation; but where they intersect, we have an opportunity to make intentional choices about our future course.
This conceptual move intentionally foregrounds the motivations for open search to highlight points of contentions and negotiations for the community to reconcile with.
We invite the open search community to engage in these critical reflections to better prepare for the inevitable tensions that all movements for openness must contend with.
\section{ACKNOWLEDGEMENTS}
The authors gratefully acknowledge thoughtful feedback from Sireesh Gururaja in the early stages of this work.
No external funding was received in support of this work.

\printbibliography
\end{document}